\documentclass{statsoc}
\usepackage{amssymb}
\usepackage{amsmath}
\usepackage{amsfonts}
\usepackage{graphics, graphicx}
\usepackage{natbib}
\newtheorem{definition}{Definition}

\newcommand{\inthh}{{\int_{-1/2}^{1/2}}}

\def\cov{\mathop{\rm cov}\nolimits}

\def\var{\mathop{\rm var}\nolimits}

\def\inthh{\int_{-1/2}^{1/2}}

\def\enbox#1{\enskip\hbox{#1}\enskip}
\newfont{\Bb}{msbm10}
\newfont{\Bbs}{msbm7}

\newcommand{\ints}{\mbox{\Bb Z}}

\newcommand{\posints}{\mbox{\Bb Z}^+}

\title[ASSESSING CHARACTERISTIC SCALES USING WAVELETS]{Assessing Characteristic Scales Using Wavelets
\thanks{ {\it Address for correspondence}: Donald B.~Percival, Applied Physics Laboratory, Box 355640, Univerisity of Washington, Seattle, WA, 98195--5640, USA. E-mail: dbp@apl.washington.edu }}
\author[Keim and Percival]{Michael J.~Keim}
\address{DNV Renewables (USA) Inc.,
Seattle, WA, USA} \email{dbp@apl.washington.edu}
\author[Keim and Percival]{Donald B.~Percival}
\address{University of Washington, Seattle, WA, USA}

\begin{document}
\begin{abstract}
Characteristic scale is a notion that pervades the geophysical sciences, but it has no widely accepted precise definition.  The wavelet transform decomposes a time series into coefficients that are associated with different scales.  The variance of these coefficients can be used to decompose the variance of the time series across different scales.  A practical definition for characteristic scale can be formulated in terms of peaks in plots of the wavelet variance versus scale.  This paper presents basic theory for characteristic scales based upon the discrete wavelet transform, proposes a natural estimator for these scales and provides a large sample theory for this estimator that permits the construction of confidence intervals for a true unknown characteristic scale.  Computer experiments are presented that demonstrate the efficacy of the large sample theory for finite sample sizes.  Examples of characteristic scale estimation are given for global temperature records, coherent structures in river flows, the Madden--Julian oscillation in an atmospheric time series and transects of one type of Arctic sea ice.
\end{abstract}

\section{Introduction}\label{sec:Intro}
Time series in geophysics and other areas
often seem to be describable as a series of `states' or `events'
whose durations tend to cluster around a value known as a characteristic scale.
Although the notion of characteristic scale is widespread in the physical sciences,
it does not have a precise definition
independent of summary statistics
that have been proposed
to extract it from particular time series.
Several definitions for characteristic scale
are discussed in von Storch and Zwiers~(1999)
for time series
that can be modeled as a stochastic process
$X_t$, $t\in\ints$
(the set of all integers).
One definition involves quantifying the `memory' of the process.
Suppose that ${\bf P}[X_{t+\tau} > 0 \,\vert\,  X_{t} > 0] > 0.5$
for small lags $\tau$,
but ${\bf P}[X_{t+\tau} > 0 \,\vert\,  X_{t} > 0] = 0.5$
at large lags.
The smallest $\tau$ such that the latter relationship holds
is one way to define a characteristic scale.
Although this definition has some intuitive appeal
because it is based on the length of time
that a process takes to `forget' its current positive state,
von Storch and Zwiers
note that it is of limited practical value;
for example,
it leads to an infinite characteristic scale
for a first-order autoregressive (AR(1)) process,
one of the most popular models in time series analysis.
Under the additional assumption that
$X_t$ is a wide-sense stationary process with variance $\sigma^2$,
a more useful definition compares $X_t$ to a process $Y_t$
consisting of independent and identically distributed random variables,
also with variance $\sigma^2$.
The sample mean of $Y_1, Y_2, \ldots, Y_N$ has variance $\sigma^2/N$,
whereas that for $X_1, X_2, \ldots, X_N$
can be expressed as $\sigma^2/N'$,
where $N'$ is referred to as the equivalent sample size.
The limit of the ratio $N/N'$ as $N\to\infty$ defines
a quantity $\tau_D$ known as the decorrelation time.
As von Storch and Zwiers argue,
$\tau_D$ is a reasonable definition for characteristic scale
for some -- but not all -- time series.
The practicality of this measure is partly
due to its relationship with the autocorrelation sequence (ACS)
$\rho_k$ for $X_t$, namely,
\begin{equation}\label{eq:decorrelation}
\tau_D = 1 + 2 \sum_{k=1}^\infty \rho_k.
\end{equation}
Appropriate estimates of $\rho_k$ can thus be used
as the basis for an estimate of $\tau_D$;
for an AR(1) process,
we have $\tau_D = (1+\rho_1)/(1-\rho_1)$.
More generally,
von Storch and Zwiers note that
the variances of statistics other than the sample mean
can be used in a similar manner 
to define other measures of characteristic scale.
Other approaches for defining characteristic scale
have been discussed in the literature.
Four examples are
Simonetti et al.~(1985),
who cast the definition in terms of the structure function
(basically a reformulation of the ACS);
Cordes~(1986), who uses the shape of the ACS;
Higuchi~(1988),
who links characteristic scale to a measure of fractal dimension; and
Tsonis et al.~(1998),
who define the concept
in terms of fluctuations from cumulative sums
in combination with detrending via singular spectrum analysis
(see Section~\ref{sec:globaltemp} for details).

In this article we propose a new definition for characteristic scale
based upon the discrete wavelet transform (DWT) of $X_t$.
The DWT is often described as a scale-based transform
(see, e.g., Flandrin, 1999, Percival and Walden, 2000,
and Nason, 2008).
To fix ideas, let us focus on the Haar DWT.
This transform yields wavelet coefficients, say $W^{\hbox{\tiny Haar}}_{\tau,t}$,
that reflect changes in adjacent averages 
spanning integer scales $\tau$
at times indexed by $t$; to be precise,
\[
W^{\hbox{\tiny Haar}}_{\tau,t}
\propto
\frac{1}{\tau} \sum_{l=0}^{\tau-1} X_{t-l}
-
\frac{1}{\tau} \sum_{l=0}^{\tau-1} X_{t-\tau-l}.
\]
If the `events' in a time series
have a characteristic duration of $\tau$,
then $|W^{\hbox{\tiny Haar}}_{\tau,t}|$
will tend to be large at certain indices $t$.
Under the assumption that $X_t$ is a stationary process,
a summary of the `largeness' of $|W^{\hbox{\tiny Haar}}_{\tau,t}|$ across $t$ 
is provided by a time-independent quantity $\var\,\{W^{\hbox{\tiny Haar}}_{\tau,t}\}$
known as the wavelet variance.
The wavelet variance provides a scale-based decomposition of the variance of $X_t$
(see Section~\ref{sec:WVDefinition} for details),
so a large $\var\,\{W^{\hbox{\tiny Haar}}_{\tau,t}\}$ for a particular $\tau$
should provide the basis for defining a characteristic scale
that is in the neighborhood of $\tau$.
The goal of this article is to expand this key idea
to define a wavelet-based characteristic scale
and to provide theory for a corresponding statistically tractable estimator.

The remainder of this paper is organized as follows.
Section~\ref{sec:WVBackground} gives some necessary background
on the DWT, the wavelet variance and its sampling theory
for intrinsically stationary Gaussian processes.
Section~\ref{sec:WBDCS} proposes a wavelet-based definition
of characteristic scale and contrasts it with $\tau_D$.
Section~\ref{sec:CSEstTheory} deals with estimation of
the wavelet-based characteristic scale
and provides some large-sample theory for the proposed estimator.
Section~\ref{sec:MCexperiments} reports on a Monte Carlo study
that examines the efficacy of the large-sample theory
for a representative selection of processes
and finite sample sizes.
Section~\ref{sec:examples} gives four examples of estimating
characteristic scales for actual time series.
The final section (\ref{sec:Summary}) has a summary and
a discussion of possible extensions.

\section{Background on the Wavelet Variance}\label{sec:WVBackground}
Here we define the wavelet variance,
give an interpretation for it
and present formulae that allows its computation,
after which we review its estimation theory.
\subsection{Definition and Basic Properties of the Wavelet Variance}\label{sec:WVDefinition}
Let $\{ X_t \}$ be an intrinsically stationary process of integer order $d\ge0$,
defined as follows.
If $d=0$, $\{ X_t \}$ itself is stationary
in the sense that
both $E\{ X_t \}$
and $\cov\,\{X_{t+\tau},X_t\}$
exist, are finite and are independent of $t$;
if $d>0$,
then subjecting $\{ X_t \}$ to a $d$th order backward difference filter
yields a stationary process, namely,
\[
X^{(d)}_t \equiv \sum_{k=0}^d {d \choose k} (-1)^k X_{t-k},
\]
whereas $\{ X^{(d-1)}_t \}, \ldots, \{ X^{(1)}_t \}$ and $\{ X^{(0)}_t \} \equiv \{ X_t \}$
are all nonstationary.
Under the assumption that $\{ X^{(d)}_t \}$ has
a spectral density function (SDF) denoted by $S_{X^{(d)}}(\cdot)$,
the (generalized) SDF for $\{ X_t \}$ is defined to be
\[
S_X(f) = \frac{S_{X^{(d)}}(f)}{[4\sin^2(\pi f)]^d},
\]
where $4\sin^2(\pi f)$ defines the squared gain function
for a first-order backward difference filter
(Yaglom, 1958).
We denote the autocovariance sequence (ACVS) for $\{ X^{(d)}_t \}$
by $\{ s^{(d)}_\tau \}$.

Let $\{ h_{1,l} : l= 0, 1, \ldots, L_1-1 \}$
be a unit-level Daubechies wavelet filter of width $L_1=2, 4, 6, \ldots$
normalized such that $\sum_l h^2_{1,l} = 1/2$
(Daubechies, 1992).
If $L_1\ge4$,
use of this filter is equivalent to subjecting the output
from a backward difference filter of order $d=L_1/2$
to a low-pass filter of width $L_1/2$
(the case $L_1=2$ yields the Haar wavelet filter,
whose coefficients $\{ \frac{1}{2}, -\frac{1}{2}\}$
are proportional to a first-order backward difference filter). 
Let $g_{1,l} \equiv (-1)^{l+1}h_{1,L_1-1-l}$
be the corresponding scaling filter.
Let
\[
H_1(f)
\equiv
\sum_{l=0}^{L_1-1} h_{1,l} e^{-i2\pi f l}
\]
define the transfer function for the wavelet filter,
and let $G_1(f)$ denote the same for the scaling filter.
For a level $j \ge 2$,
let 
\[
H_j(f)
\equiv
H_1(2^{j-1}f) \prod_{k=0}^{j-2} G_1(2^kf).
\]
The inverse Fourier transform of this function
gives the impulse response sequence
for the $j$th level wavelet filter
$\{ h_{j,l}: l=0,\ldots L_j-1\}$,
where $L_j \equiv (2^j-1)(L_1-1)+1$.
We denote the corresponding squared gain function by
$
{\cal H}_j(f) =
| H_j(f) |^2.
$
The filter $\{ h_{j,l} \}$ is approximately a bandpass filter
with a passband given by $|f|\in(1/2^{j+1}, 1/2^j]$.

The $j$th level wavelet coefficient process for $\{ X_t \}$ is given by
\[
W_{j,t}
\equiv
\sum_{l=0}^{L_j-1} h_{j,l} X_{t-l}.
\]
The coefficient $W_{j,t}$
is proportional to changes in adjacent weighted averages
with an effective scale (or span) of $\tau_j=2^{j-1}$.
Note that scale $\tau_j$ is associated with the frequency interval
$(1/(4\tau_j), 1/(2\tau_j)]$ and the interval of periods $[2\tau_j, 4\tau_j)$.
Under the assumptions that
$\{ X_t \}$ is an intrinsically stationary process of order $d$
with SDF $S_X(\cdot)$
and that $L_1\ge 2d$,
$\{ W_{j,t} \}$ is a stationary process with SDF given by
\[
S_j(f)
\equiv
{\cal H}_j(f) S_X(f)
= 
\frac{{\cal H}_j(f) S_{X^{(d)}}(f)}{[4\sin^2(\pi f)]^d}.
\]
By definition the wavelet variance
for $\{ X_t \}$ at scale $\tau_j$
is the variance of $\{ W_{j,t} \}$:
\[
\nu^2_j \equiv \var\,\{ W_{j,t} \}
= \inthh S_j(f) \,df
= \inthh {\cal H}_j(f) S_X(f)\,df.
\]
If $\{ X_t \}$ is a stationary process, then
\[
\var\,\{ X_t \}
= \sum_{j=1}^\infty \nu^2_j
\]
(Percival, 1995),
and the wavelet variance for scale $\tau_j$
can be interpreted
as the contribution to the overall variance
due to changes in adjacent weighted averages over that scale
(if $\{ X_t \}$ is nonstationary, the summation above
diverges to infinity,
but $\nu^2_j$ still has the interpretation
of measuring the variability of changes
in adjacent weighted averages).

The theoretical wavelet variance for
an intrinsically stationary process can be computed
readily in terms of the ACVS $\{ s^{(d)}_{\tau} \}$
for its underlying stationary component.
In particular, we can write
\[
\nu^2_j
=
s^{(d)}_{0} \sum_{l=0}^{L_j-d-1} \left(b^{(d)}_{j,l}\right)^2
+
2\sum_{\tau=1}^{L_j-d-1}
s^{(d)}_{\tau}
\sum_{l=0}^{L_j-d-1-\tau} b^{(d)}_{j,l} b^{(d)}_{j,l+\tau},
\]
where $\{ b^{(d)}_{j,l} \}$ is the $d$th-order cumulative summation
of $\{ h_{j,l} \}$; i.e., with $b^{(0)}_{j,l} \equiv h_{j,l}$,
we have, for $k=1,\ldots,d$,
\[
b^{(k)}_{j,l} = \sum_{n=0}^l b^{(k-1)}_{j,n},
\quad
l=0, 1, \ldots, L_j - k - 1
\]
(Lemma 1, Craigmile and Percival, 2005).
Using $\{ b^{(d)}_{j,l} \}$,
we can write
\[
W_{j,t}
=
\sum_{l=0}^{L_j-d-1} b^{(d)}_{j,l} X^{(d)}_{t-l}.
\]
Denote the transfer function and squared gain function
for $\{ b^{(d)}_{j,l} \}$ as
\[
B^{(d)}_j(f) \equiv
\sum_{l=0}^{L_j-d-1} b^{(d)}_{j,l} e^{-i2\pi f l}
\enbox{and}
{\cal B}^{(d)}_j(f) \equiv
| B^{(d)}_j(f) |^2 =
\frac{{\cal H}_j(f)}{[4\sin^2(\pi f)]^d}.
\]
Then we can have
\[
S_j(f) =  {\cal B}^{(d)}_j(f) S_{X^{(d)}}(f)
\enbox{and hence}
\nu^2_j 
= \inthh {\cal B}^{(d)}_j(f) S_{X^{(d)}}(f) \,df.
\]

\subsection{Estimation Theory for the Wavelet Variance}\label{sec:WVEstTheory}

Given a time series  that can be regarded as a realization of a portion
$X_0, X_1, \ldots, X_{N-1}$ of length $N$
of the process $\{ X_t \}$,
we can compute the level $j$ wavelet coefficients
for indices $L_j -1 \le t \le N-1$
under the assumption that $M_j \equiv N - L_j +1 > 0$.
A sufficient (but not necessary) condition for $\{ W_{j,t} \}$
to be a zero mean stationary process
is that $L_1 > 2d$
(if $L_1 = 2d$,
then $\{ W_{j,t} \}$
is necessarily stationary, but it might not have zero mean).
Assuming that $L_1$ is chosen such
that $\{ W_{j,t} \}$
is a zero mean stationary process,
we have
$
\nu^2_j = E\{ W^2_{j,t} \}
$
and hence
\begin{equation}\label{eq:unbiasedWVest}
\hat \nu^2_j
\equiv
\frac{1}{M_j}
\sum_{t=L_j-1}^{N-1}
W^2_{j,t}
\end{equation}
is an unbiased estimator of the wavelet variance.

To look at the second moment properties of $\hat \nu^2_j$,
we assume that the $W_{j,t}$ obey a multivariate Gaussian distribution.
Using the Isserlis theorem (Isserlis, 1918)
and assuming $j\le k$,
we find that variance and covariance of
$\hat \nu^2_j$ and $\hat \nu^2_k$
are given by
\begin{equation}\label{eq:covWVjWVk}
\cov\,\{ \hat \nu^2_j, \hat \nu^2_k \}
=
\frac{2}{M_j}
\sum_{\tau=-(M_k-1)}^{M_k-1}
\left(1 - \frac{|\tau|}{M_k}\right)
s^2_{j,k,\tau}
+
\frac{2}{M_j M_k}
\sum_{t=L_j-1}^{L_k-2}
\sum_{u=L_k-1}^{N-1}
s^2_{j,k,t-u},
\end{equation}
where $\{ s_{j,k,\tau} \}$
is the cross-covariance sequence
for the bivariate stationary processes
$\{ W_{j,t} \}$ and
$\{ W_{k,t} \}$:
\[
s_{j,k,\tau}
\equiv
\cov\left\{
W_{j,t+\tau},
W_{k,t}
\right\}
=
\sum_{l=0}^{L_j-d-1} b^{(d)}_{j,l} 
\sum_{m=0}^{L_k-d-1} b^{(d)}_{k,m}
s^{(d)}_{\tau-l+m}
\]
(when using~(\ref{eq:covWVjWVk})
to compute $\var\,\{\hat \nu^2_j\}$ by letting $k=j$,
the double summation is interpreted as zero). 

While (\ref{eq:covWVjWVk}) is an exact result,
it is of interest to explore an approximation
to $\cov\,\{ \hat \nu^2_j, \hat \nu^2_k \}$
that leads to a practical scheme for estimating it.
As $N\to\infty$ and hence $M_k\to\infty$ also,
the double summation in (\ref{eq:covWVjWVk}) becomes negligible,
whereas the first summation can be approximated
using a Ces{\`a}ro sum argument,
which, followed by an appeal to Parseval's theorem, yields
\begin{equation}\label{eq:covWVjWVkApproxOne}
\cov\,\{ \hat \nu^2_j, \hat \nu^2_k \}
\approx
\frac{2}{M_j}
\sum_{\tau=-\infty}^\infty
s^2_{j,k,\tau}
=
\frac{2A_{j,k}}{M_j},
\enbox{where}
A_{j,k}
\equiv
\inthh S_j(f) S_k(f) \,df.
\end{equation}
Suppose that $\hat S_j(f)$, $0 < f < 1/2$,
is some standard nonparametric SDF estimator
whose large-sample distribution is dictated by $S_j(f) \chi^2_{\eta}/\eta$,
where $\chi^2_{\eta}$ is a chi-square random variable
with $\eta$ degrees of freedom (see, e.g., Priestley, 1981).
Letting $\hat S_k(f)$ be a similar estimator for $S_k(f)$,
it follows from standard theory for multivariate SDF estimation (Priestley, 1981)
that
\[
E\{ \hat S_j(f) \hat S_k(f) \} \approx S_j(f) S_k(f) \left( \frac{2}{\eta} + 1 \right)
\]
and hence that
\[
\hat A_{j,k}
\equiv
\frac{\eta}{2+\eta}
\inthh \hat S_j(f) \hat S_k(f) df
\]
is an approximately unbiased estimator of $A_{j,k}$.
Specializing to the case where
$\hat S_j(f)$ and $\hat S_k(f)$ are lag window estimators
based upon lag windows $\{ w_{j,\tau} \}$ and $\{ w_{k,\tau} \}$ (Priestley, 1981),
we have 
\begin{equation}\label{eq:AjkEst}
\hat A_{j,k}
=
\frac{\eta}{2+\eta}
\left(
\hat \nu^2_j \hat \nu^2_k
+
2
\sum_{\tau=1}^{M_k-1}
w_{j,\tau} \hat s_{j,\tau} w_{k,\tau} \hat s_{k,\tau}
\right),
\end{equation}
where $\{ \hat s_{j,\tau} \}$
is the biased estimator of the ACVS for
$W_{j,L_j-1}$, \dots, $W_{j,N-1}$:
\[
\hat s_{j,\tau}
\equiv
\frac{1}{M_j}
\sum_{t=L_j-1}^{N-1-\tau} W_{j,t+\tau} W_{j,t},
\qquad
0 \le \tau \le M_j -1.
\]
A practical scheme for approximating
$\cov\,\{\hat \nu^2_j, \hat \nu^2_k \}$
is to substitute $\hat A_{j,k}$ for $A_{j,k}$ in (\ref{eq:covWVjWVkApproxOne}).

Finally, we note that,
under a Gaussian assumption on $\{ X_t \}$
and mild conditions on its SDF,
$\hat \nu^2_j$
is asymptotically normally distributed with mean $\nu^2_j$
and large sample variance $2A_{j,j}/M_j$
(Percival, 1995; Mondal, 2007;
see Serrouk et al., 2000, for related results
that relax the Gaussian assumption).

\section{Wavelet-Based Definition of Characteristic Scale}\label{sec:WBDCS}
Because we can interpret the wavelet variance at a particular scale $\tau_j$
as the contribution to the overall variance of $\{ X_t \}$
due to changes in adjacent weighted averages over that scale,
we can formulate a wavelet-based notion of characteristic scale
by searching for scales at which $\nu^2_j$ is large compared
to its surrounding values,
thus leading to the following definitions.
\begin{definition}
(Local characteristic scale $\tau_{c,j}$.)
Suppose $\{ X_t \}$ is an intrinsically stationary process
such that $\nu^2_j \ge \nu^2_{j\pm1}$ for some $j\ge2$,
with strict inequality holding in at least one case.
Fit a quadratic $y_k = a + bx_k + c x_k^2$
that passes through $(x_k,y_k) \equiv (\log_2(\tau_k),\log_2(\nu^2_k))$,
$k=j-1, j, j+1$.
A local characteristic scale is
the location at which the quadratic is maximized:
\begin{equation}\label{eq;DefTaucj}
\tau_{c,j}
= 2^{-b/(2c)}
= 2^{-\beta_1/\beta_2} \tau_j ,
\enbox{where}
\beta_1 \equiv \frac{y_{j+1} - y_{j-1}}{2}
\enbox{and}
\beta_2 \equiv y_{j+1} - 2 y_j + y_{j-1}.
\end{equation}
\end{definition}
We note in passing that
$\tau_j/\surd2 \le \tau_{c,j} \le \tau_j\surd2$
and that the pattern $\nu^2_{j-1} < \nu^2_j = \nu^2_{j+1} > \nu^2_{j+2}$
yields $\tau_{c,j} = \tau_{c,j+1} = \tau_j\surd 2$.
\begin{definition}
(Global characteristic scale $\tau_c$.)
Suppose $\{ X_t \}$ has a local characteristic scale $\tau_{c,j}$
such that $\nu^2_j > \nu^2_k$ for all
$k\in\posints$ excluding $k=j-1,j,j+1$, 
where $\posints$ is the set of positive integers.
Then $\{ X_t \}$ is said to have a global characteristic scale $\tau_c \equiv \tau_{c,j}$.
\end{definition}
Figure~1 shows four examples of theoretical wavelet variance curves
with local characteristic scales.
If we assume that the wavelet variances at scales not depicted in the plots
are all smaller than the ones shown,
the processes associated with~(a), (b) and~(d)
have a global characteristic scale,
but the one for (c)~does not.
\begin{figure}[h]
{\includegraphics[angle=0]{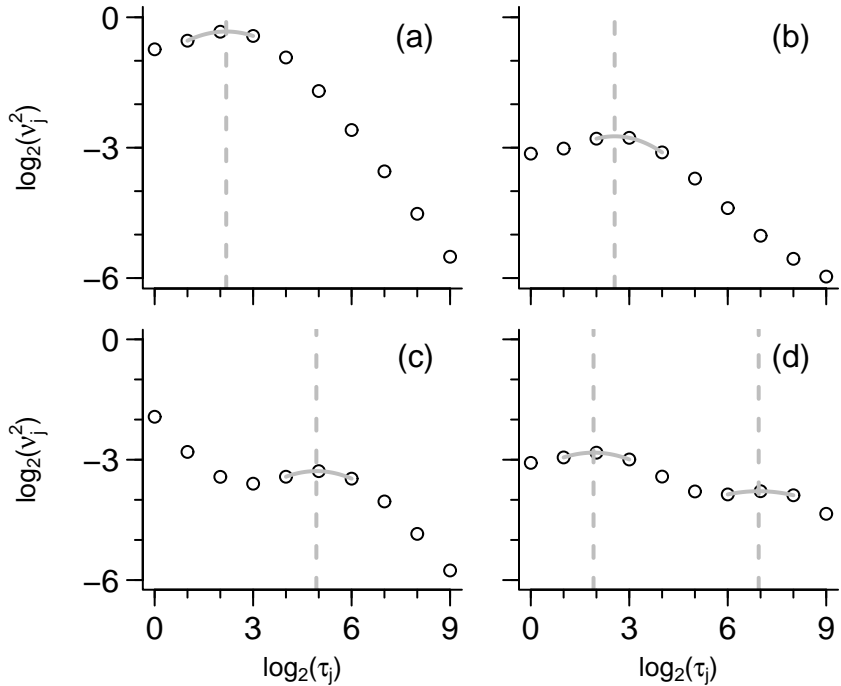}}
\vspace{-0.2in}
\caption{
Log of wavelet variances $\nu^2_j$ versus log of $\tau_j$
(circles) for
(a)~a first-order autoregressive (AR(1)) process,
(b)~a linear combination of an AR(1) process
and a fractionally differenced process;
(c)~a linear combination of an AR(1) process 
and white noise; and
(d)~a linear combination of two AR(1) processes
(see Section~\ref{sec:MCexperiments}
for precise definitions of all four processes).
The vertical dashed lines indicate
the locations the characteristic scales $\tau_{c,j}$,
while the gray curves show the quadratic fit
whose maximum location determines $\tau_{c,j}$.
}.
\label{fig:ExampleWVcurves}
\end{figure}

\begin{figure}
{\includegraphics[angle=0]{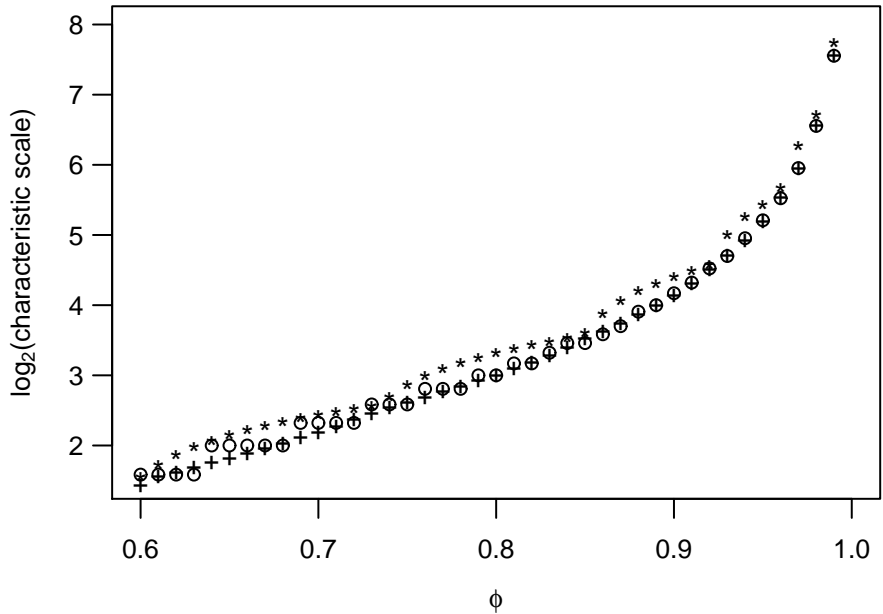}}
\vspace{-0.2in}
\caption{Comparison of three wavelet-based measures of characteristic scale
for an AR(1) process with a unit-lag autocorrelation of $\phi$.
The pluses show $\tau_{c}$,
which is based on a quadratic fit in log/log space
around the peak in the Haar wavelet variance curve
evaluated over dyadic scales $\tau_j=2^{j-1}$.
The asterisks show a similar measure,
but now based on a quadratic fit in linear/linear space.
The circles are based on a measure
given by the location of the peak of the Haar wavelet variance curve
evaluated over all integer-valued scales.
}.
\label{fig:CompareTauCdefs}
\end{figure}

Our definitions for $\tau_{c,j}$ and $\tau_{c}$ need some justification.
Arguably a more natural definition for characteristic scale
would involve a wavelet variance
defined over a continuum of scales
via a continuous wavelet transform (CWT).
A local maxima of a CWT-based wavelet variance curve
would then define a local characteristic scale,
which would seem to be preferable to our interpolation scheme
based on just the dyadic scales $\tau_j = 2^{j-1}$.
The following example suggests the CWT- and DWT-based definitions are quite similar,
which leads us to prefer the latter
because it is much easier to compute
and because its estimator is statistically tractable.
Consider an AR(1) process $X_t = \phi X_{t-1} + \epsilon_t$,
where $\{ \epsilon_t \}$ is a Gaussian white noise process with zero mean
and unit variance.
For the Haar DWT,
this process has a global characteristic scale when $0.57 < \phi < 1$.
The pluses in Fig.~\ref{fig:CompareTauCdefs} show how $\tau_{c}$
increases as $\phi$ varies from $0.60$ to $0.99$ in steps of $0.01$.
Avoiding interpolation issues
that arise in using a CWT with a time series sampled over the integers,
we can readily extend the definition of the Haar wavelet variance
to all scales $\tau\in\posints$:
\[
\nu^2(\tau)
=
\frac{\var\left\{
\sum_{l=0}^{\tau-1} X_{t-l}
-
\sum_{l=0}^{\tau-1} X_{t-\tau-l}
\right\}}
{4\tau^2}.
\]
A plot of $\nu^2(\tau)$ versus $\tau$
for $\phi = 0.60, 0.61, \ldots, 0.99$,
shows a unique maximum,
which provides us
with a CWT-like definition of characteristic scale $\tilde \tau_c\in\posints$.
The circles in Fig.~\ref{fig:CompareTauCdefs} show $\tilde \tau_{c}$ versus $\phi$.
The agreement between $\tau_{c}$ and $\tilde \tau_{c}$ is very good
and gets better as $\phi$ increases.
By contrast,
if we were to define $\tau_{c}$
in terms of a quadratic fit in linear/linear rather than log/log space,
we obtain the asterisks shown in the figure,
which do not agree nearly as well with the CWT-based definition $\tilde \tau_c$.
Use of linear/log space or log/linear space yields
characteristic scales
that are nearly identical to those obtained using, respectively,
linear/linear space or log/log space.
The choice of log/log space over log/linear space
is dictated by the fact that the log transform
acts as a variance-stabilizing transform for wavelet variance estimators
(this can be seen further on by noting that,
while the elements of $\Sigma_1$ in Equation~(\ref{eq:CovApproxOne})
depend on the wavelet variance,
this quantity ratios out
in the elements of $\Sigma_2$ in Equation~(\ref{eq:CovApproxTwo})).
\begin{figure}
{\includegraphics[angle=0]{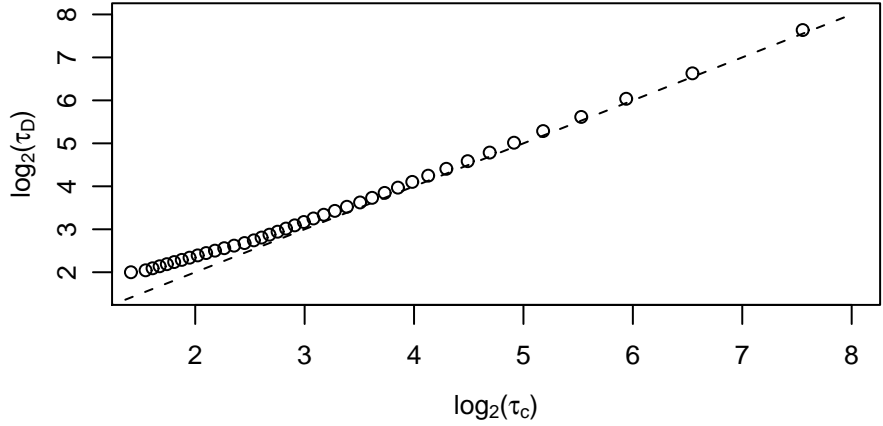}}
\vspace{-0.2in}
\caption{Decorrelation time $\tau_D$ versus characteristic scale $\tau_c$
for AR(1) processes with unit-lag autocorrelations of
$\phi = 0.60, 0.61,...,0.99$ (circles from left to right).
If $\tau_D$ and $\tau_c$ had been equal,
the circles would have fallen on the dashed line.
}.
\label{fig:CompareTauCandTauD}
\end{figure}

For a stationary process $\{ X_t \}$
with ACS $\{ \rho_k \}$,
it is of interest to compare $\tau_c$
to the measure of characteristic scale provided by the decorrelation time
of Equation~(\ref{eq:decorrelation}).
For an AR(1) process,
$\rho_k = \phi^{|k|}$ decays exponentially,
and we have $\tau_D = (1+\phi)/(1-\phi)$.
Figure~\ref{fig:CompareTauCandTauD} shows $\tau_D$ versus the Haar-based $\tau_c$
as $\phi$ ranges over $0.60, 0.61, \ldots, 0.99$.
The two measures track each other as $\phi$ gets large,
with $\tau_D \doteq 1.01 \tau_c$ for $\phi=0.99$.
Figure~1(a) shows the Haar wavelet variance versus $\tau_j$
when $\phi = 0.7$, for which $\tau_c \doteq 4.53$.
By contrast, Fig.~1(b) shows a similar wavelet variance curve
for a process that is a linear combination
of an AR(1) process with $\phi=0.75$
and a fractionally differenced (FD) process with long-memory parameter $\delta = 0.45$
(Granger and Joyeux, 1980; Hosking, 1981; Beran, 1984).
The two processes are independent of each other.
Here $\tau_c \doteq 5.87$,
whereas $\tau_D$ is infinite
because $\rho_k$ decays hyperbolically
due to the influence of the FD process.
This fact points out a fundamental difference
between the measures $\tau_c$ and $\tau_D$:
whereas the former concentrates on localized properties of $\{ X_t \}$,
the latter is influenced to a large degree
by the asymptotic decay rate of the ACS
(and cannot be used at all with intrinsically stationary processes
with $d>0$).

\section{Statistical Properties of Characteristic Scale Estimators}\label{sec:CSEstTheory}
Suppose we have a time series that can be regarded as a realization
of a portion $X_0$, $X_1$, \ldots, $X_{N-1}$ of an intrinsically stationary process
of order $d$,
based upon which we want to estimate local characteristic scales (assuming such exist)
for $\{ X_t \}$.
We start by estimating the wavelet variance
out to some maximum scale of interest $\tau_{J_0}$
using the unbiased estimator
$\hat \nu^2_j, j = 1, \ldots, J_0$.
If there is some $1 < j < J_0$ such that the estimates
obey the pattern $\hat \nu^2_j \ge \hat \nu^2_{j\pm1}$
(with strict inequality holding in at least one case),
we can define an estimator $\hat \tau_{c,j}$ of the characteristic scale
in the region of $\tau_{j}$
by replacing $y_k$ with $\hat y_k \equiv \log_2(\hat \nu^2_k)$
in equation~(\ref{eq;DefTaucj}):
\[
\hat \tau_{c,j} 
=
2^{- \hat\beta_1/\hat\beta_2}\tau_j,
\enbox{where}
\hat\beta_1 \equiv \frac{\hat y_{j+1} - \hat y_{j-1}}{2}
\enbox{and}
\hat\beta_2 \equiv \hat y_{j+1} - 2\hat y_j + \hat y_{j-1}.
\]
We want to establish simple -- but reasonable -- approximations
to the sampling properties of this estimator
under the assumption that $N$ is large.

Motivated by the large-sample theory reviewed at the end of Section~\ref{sec:WVEstTheory},
we start with the assumption that
$
\left[ \hat \nu^2_{j-1}, \hat \nu^2_j, \hat \nu^2_{j+1} \right]^T
$
is multivariate Gaussian with a mean given by
the true wavelet variances
$
\left[ \nu^2_{j-1}, \nu^2_j, \nu^2_{j+1} \right]^T
$
and a covariance matrix dictated by equation~(\ref{eq:covWVjWVk}).
We use equation~(\ref{eq:covWVjWVkApproxOne})
to approximate this symmetric matrix by
\begin{equation}\label{eq:CovApproxOne}
\Sigma_1
\equiv
2\left[
\begin{array}{ccc}
{A_{j-1,j-1}}/{M_{j-1}}&\hbox{---}&\hbox{---}\\
{A_{j-1,j}}/{M_{j-1}}&{A_{j,j}}/{M_{j}}&\hbox{---}\\
{A_{j-1,j+1}}/{M_{j-1}}&{A_{j,j-1}}/{M_{j}}&{A_{j+1,j+1}}/{M_{j+1}}\\
\end{array}
\right].
\end{equation}
The delta method says that 
$
\left[ \log_2\left(\hat \nu^2_{j-1}\right), \log_2\left(\hat \nu^2_j\right), \log_2\left(\hat \nu^2_{j+1}\right) \right]^T
= \left[ \hat y_{j-1}, \hat y_j, \hat y_{j+1} \right]^T
$
is approximately Gaussian with mean
$
\left[ \log_2\left(\nu^2_{j-1}\right), \log_2\left(\nu^2_j\right), \log_2\left(\nu^2_{j+1}\right) \right]^T
=
\left[
y_{j-1}, y_j, y_{j+1}
\right]^T
$
and covariance matrix $\Sigma_2$ whose elements are
\begin{equation}\label{eq:CovApproxTwo}
\Sigma_{2,m,n}
\equiv
\frac{\cov\,\{ \hat \nu^2_{j-2+m}, \hat \nu^2_{j-2+n}\}}{\nu^2_{j-2+m}\nu^2_{j-2+n} \log^2(2)}
+
2
\frac{\var\,\{ \hat \nu^2_{j-2+m} \} \var\,\{ \hat \nu^2_{j-2+n} \} +
(\cov\,\{\hat \nu^2_{j-2+m},\hat \nu^2_{j-2+n}\})^2}{\nu^4_{j-2+m}\nu^4_{j-2+n} \log^2(2)},
\end{equation}
with $m$ and $n=1,2$ and 3.
Since 
\[
\left[
\begin{array}{c}
\hat\beta_1\\
\hat\beta_2
\end{array}
\right]
=
\left[
\begin{array}{ccc}
-\frac{1}{2}&0&\frac{1}{2}\\
1&-2&1
\end{array}
\right]
\left[
\begin{array}{c}
\hat y_{j-1}\\
\hat y_{j}\\
\hat y_{j+1}
\end{array}
\right]
\equiv
H
\left[
\begin{array}{c}
\hat y_{j-1}\\
\hat y_{j}\\
\hat y_{j+1}
\end{array}
\right],
\]
it follows that $[\hat\beta_1,\hat\beta_2]^T$ is approximately Gaussian with
mean $[\beta_1,\beta_2]^T$
and covariance $H\Sigma_2 H^T$.
Let $\hat \kappa \equiv - \hat\beta_1/\hat\beta_2$.
Further applications of the delta method say that
$\hat \kappa$ is approximately Gaussian with mean
$-\beta_1/\beta_2$ and variance given approximately by
\begin{eqnarray}
\sigma^2_{\hat \kappa}
&\equiv&
\frac{\var\,\{ \hat \beta_1 \}}{\beta^2_2}
+ \frac{\beta^2_1 \var\,\{ \hat \beta_2 \}}{\beta^4_2}
+ \frac{\var\,\{ \hat \beta_1 \} \var\,\{ \hat \beta_2 \}
+ 2(\cov\,\{\hat \beta_1, \hat \beta_2\})^2} {\beta^4_2}\nonumber\\
\label{eq:varKappa}
&&
+ \frac{3 \beta^2_1 (\var\,\{ \hat \beta_2 \} )^2}{\beta^6_2}
-  \frac{2\beta_1\cov\,\{\hat \beta_1, \hat \beta_2\}}{\beta_2^3}
\end{eqnarray}
and that
\[
\var\,\{ \hat \tau_{c,j}\}
\approx
\tau^2_{c,j} \,\sigma^2_{\hat \kappa}\,\log^2_e(2).
\]
We can now provide, for example,
an approximate 95\% confidence interval (CI) $[L_-,L_+]$ for $\kappa$;
i.e.,
\[
{\bf P}[ L_- \le \kappa \le L_+ ] \approx 0.95,
\enbox{where}
L_\pm \doteq \hat \kappa \pm 1.96 \sigma_{\hat \kappa}.
\]
The event $L_- \le \kappa \le L_+$ is equivalent to the event
$\tau_j 2^{L_-} \le \tau_{c,j} \le \tau_j 2^{L_+}$,
so an approximate 95\% CI for 
$\tau_{c,j}$ is given by $[2^{-1.96 \sigma_{\hat \kappa}} \hat \tau_{c,j}, 2^{1.96 \sigma_{\hat \kappa}} \hat \tau_{c,j}]$.
In practical applications,
we can estimate $\sigma^2_{\hat \kappa}$
in a `plug-in' manner
by using $\hat A_{j,k}$ from~(\ref{eq:AjkEst})
for $A_{j,k}$ in (\ref{eq:CovApproxOne}),
$\hat \nu^2_j$ from~(\ref{eq:unbiasedWVest}) for $\nu^2_j$
in~(\ref{eq:CovApproxTwo}),
and 
$\hat \beta_1$ and $\hat \beta_2$
for $\beta_1$ and $\beta_2$ in~(\ref{eq:varKappa}).

A caveat about our approach is that it is conditioned
upon the observed pattern $\hat \nu^2_j \ge \hat \nu^2_{j\pm1}$
correctly indicating the presence of a local characteristic scale
somewhere in the vicinity of $\tau_j$.
As $N\to\infty$,
observed patterns will agree better and better
with true patterns
because of the asymptotic properties
of the wavelet variance estimators,
but observed patterns might be deceptive for finite sample sizes.
A sanity check that sheds some light on the validity
of an observed pattern
is to generate a large number of independent realizations
from a trivariate Gaussian distribution
with mean vector $\left[ \hat \nu^2_{j-1}, \hat \nu^2_j, \hat \nu^2_{j+1} \right]^T$
and covariance matrix
dictated by~(\ref{eq:CovApproxOne})
with $A_{j,k}$ replaced by $\hat A_{j,k}$ of equation~(\ref{eq:AjkEst}).
If the proportion of realizations
that have a maximum in the same location as the observed pattern
is large,
then we have some reassurance
that the observed pattern is faithfully mimicking the true pattern
(see Section~\ref{sec:ThirdSeries} for an example of this procedure).
\begin{table}
\bigskip
\caption{\label{tab:MonteCarloHaar}Results from Monte Carlo experiments (see text for details).
}
\fbox{%
\begin{tabular}{c|c|c|c|c|c}
process&$N$&$\tau_c$&$\hat\tau_c$&$M$&\% coverage\\
\hline\hline
(a)& 512&4.53&4.69& 992&88.2\\
   &2048&    &4.66&1000&87.1\\
   &8192&    &4.57&1000&94.4\\
\hline
(b)& 512&5.87&6.16& 981&89.2\\
   &2048&    &5.84&1000&90.5\\
   &8192&    &5.83&1000&93.8\\
\hline
(c)& 512&30.42&33.51& 838&86.8\\
   &2048&     &32.49& 964&88.5\\
   &8192&     &31.41&1000&93.1\\
\hline
(d)& 512&3.76&4.00& 971&91.0\\
   &2048&    &3.84& 999&96.7\\
   &8192&    &3.76&1000&96.1\\
\hline
(d)& 512&122.96& 89.31&454&86.8\\
   &2048&      &149.60&703&88.9\\
   &8192&      &153.43&775&86.7\\
\end{tabular}}
\end{table}
\section{Monte Carlo Experiments}\label{sec:MCexperiments}
We consider the following four zero-mean Gaussian stationary processes,
whose wavelet variances are depicted in Fig.~\ref{fig:ExampleWVcurves}:
\begin{enumerate}
\item
an AR(1) process
with a variance of 4 and a unit-lag autocorrelation of $\phi=0.7$;
\item
a process given by $\frac{\surd2}{\surd3}X_t + \frac{1}{\surd3}Y_t$,
where $\{ X_t \}$ is an AR(1) process with $\phi=0.75$,
while $\{ Y_t \}$ is an FD  process
with  long-memory parameter $\delta=0.45$;
\item
$\frac{1}{\surd2}X_t + \frac{1}{\surd2}Y_t$,
where $\{ X_t \}$ is an AR(1) process with $\phi=0.95$,
while $\{ Y_t \}$ is a white noise process; and
\item
$\frac{\surd2}{\surd3}X_t + \frac{1}{\surd3}Y_t$,
where $\{ X_t \}$ is an AR(1) process with $\phi=0.65$,
while $\{ Y_t \}$ is a similar process with $\phi=0.99$.
\end{enumerate}
For creating the last three processes, 
$\{ X_t \}$ and $\{ Y_t \}$
are unit variance Gaussian processes such that 
$X_s$ and $Y_t$ are independent
for all $s$ and $t$.

For each process
and for samples size $N=512$, $2048$ and $8192$,
we generated 1000 realizations 
using an exact simulation method
for AR processes
(Kay, 1981)
and for FD processes
(Davies and Harte, 1987; Craigmile, 2003). 
We recorded the number of replications $M$
for which there was a peak in the Haar wavelet variance curve
at either the proper level $j$ or
levels $j\pm1$.
For each of these $M$ realizations,
we estimated the characteristic scale
and computed a 95\% CI
using the plug-in procedure described above
(the estimates $\hat A_{j,k}$ were formed using periodograms,
which are a special case of a lag window estimator with
$w_{j,\tau} = 1$ for all $\tau$ and with $\eta=2$).
Table~\ref{tab:MonteCarloHaar}
shows the average of the estimated characteristic scales
and the percentage of times
that the 95\% CIs trapped the true characteristic scale.
There is a tendency for $\hat \tau_c$ to be positively biased.
The closeness of coverage percentage to the nominal 95\%
tends to depend upon the true $\tau_c$:
the smaller $\tau_c$ is, the better the coverage rate.
For small sample sizes, the coverage rate tends to be below the nominal 95\%.
The coverage rates tend to improve with increasing sample size,
as asymptotic theory would suggest. 
These experiments show that
the large-sample theory gives useful -- but admittedly not perfect -- approximations
to the variability in $\hat \tau_c$ for moderate sample sizes.
(Similar results were obtained using Daubechies wavelet filters
of lengths $L_1=4$ and $8$.) 
\begin{figure}
{\includegraphics[angle=0]{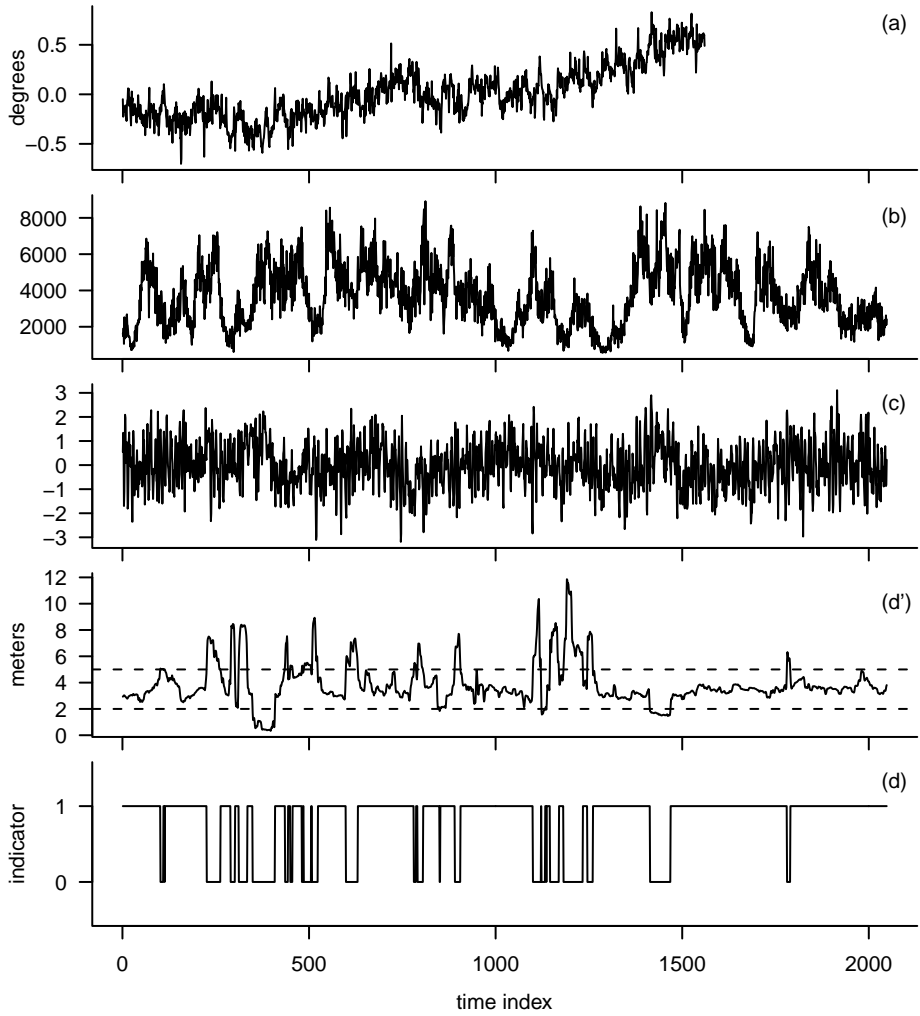}}
\caption{Four time series (upper four plots) and one derived series (bottom plot).
The series are
(a)~monthly global temperature anomalies;
(b)~coherent structures in river flows;
(c)~200-hPa velocity potential anomalies in the atmosphere;
(d')~Arctic sea ice thickness; and
(d)~indicator series for medium multiyear Arctic sea ice.
}
\label{fig:figFiveTimeSeries}
\end{figure}

\section{Real-World Examples}\label{sec:examples}
Here we consider four examples of characteristic scale estimation
based upon actual time series and the Haar wavelet.

\subsection{Global Temperature Record}\label{sec:globaltemp} 
Figure~\ref{fig:figFiveTimeSeries}(a)
shows a time series of length $N=1560$
of monthly global temperature anomalies
(land and ocean combined)
from January 1880 up to December 2009
(data obtained from {\tt http://www.ncdc.noaa.gov/pub/data/anomalies/}).
The series appears to have a prominent trend upwards.
Tsonis et al.~(1998) analyzed a closely related series
by forming cumulative sums and considering
standard deviations of lagged differences of these sums
as a function of lag.
Because trends adversely affect this method,
they elected to detrend the series
using the first three empirical orthogonal functions
extracted from a singular spectrum analysis (Elsner and Tsonis, 1996).
Subtraction of these functions from the time series
yielded a residual series
that was subjected to a test for non-zero trend
using a Cox--Stuart nonparametric test (Cox and Stuart, 1955).
Since the test failed to reject the null hypothesis of no trend,
they used the residuals to estimate a characteristic scale,
obtaining a value of about $20$~months.
This scale was the value at which the curve of standard deviations
versus lag exhibited a change in slope in log/log space.
They interpreted this characteristic scale as being
due to the influence of El Ni{\~n}o/La Ni{\~n}a cycles
on global temperatures.

\begin{figure}[h]
{\includegraphics[angle=0]{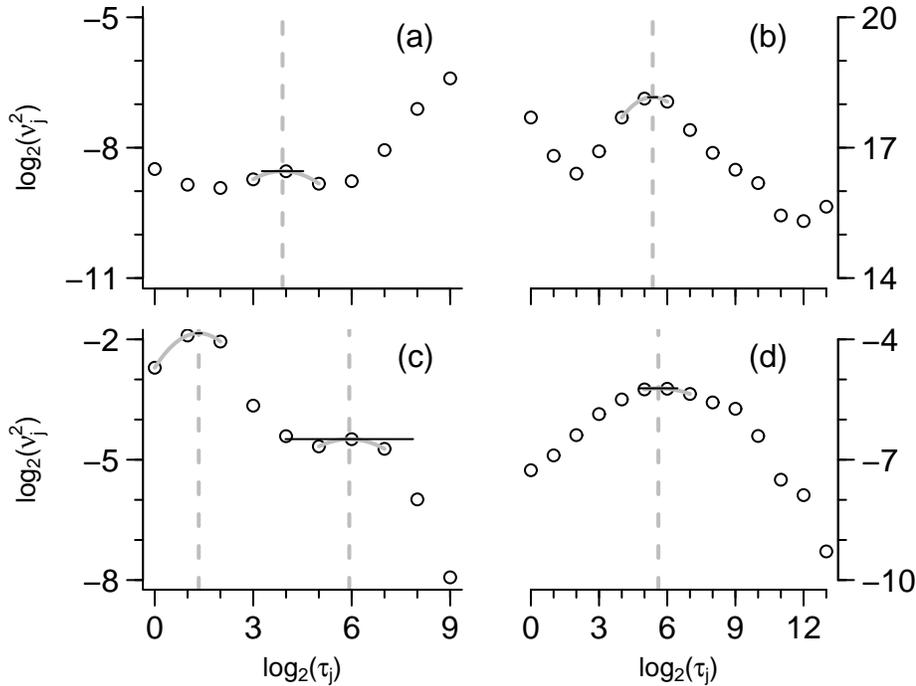}}
\vspace{-0.2in}
\caption{
Log of wavelet variances $\nu^2_j$ versus log of $\tau_j$
(circles) for
(a)~monthly global temperature anomalies;
(b)~coherent structures in river flows;
(c)~200-hPa velocity potential anomalies; and
(d)~medium multiyear Arctic sea ice.
The vertical dashed lines indicate
the locations the estimated characteristic scales $\hat \tau_{c,j}$,
while the gray curves show the quadratic fit
whose maximum location determines $\hat\tau_{c,j}$.
The horizontal solid lines depict 95\% confidence intervals
for the true characteristic scales.
}.
\label{fig:FourTimeSeriesWVcurves}
\end{figure}

Figure~\ref{fig:FourTimeSeriesWVcurves}(a) shows
that the Haar wavelet variance curve for this series
has a peak at scale $\tau_5 = 16$~months.
The corresponding estimated characteristic scale
is $\hat \tau_{c,5} \doteq 14.9$~months,
with an associated 95\% CI of $[9.6,23.0]$.
This interval traps the nominal characteristic scale
found by Tsonis et al.~(1998).
There is, however, some cause for concern here
due to the apparent trend in this time series.
If we model the series as $X_t = a + bt + Y_t$,
where $a+bt$ describes a linear (first-order polynomial) trend
and $\{ Y_t\}$ is a zero mean stationary process,
then the Haar wavelet coefficient process $\{ W_{j,t} \}$
is stationary, but has a nonzero mean,
which is in conflict with the zero mean assumption
used to construct $\hat \nu^2_j$ of Equation~(\ref{eq:unbiasedWVest}).
The wavelet coefficients most influenced by a linear
trend are those at the largest scales,
which explains the upward pattern in the wavelet variance curve
of Fig.~\ref{fig:FourTimeSeriesWVcurves}(a) at those scales.
Since there is some concern that the wavelet variance estimates
at smaller scales might also be adversely affected by the trend,
we also considered a wavelet variance curve constructed
using a Daubechies wavelet filter of width $L_1=8$
(the so-called `least asymmetric' filter).
This filter is capable of completely eliminating a trend
that is well-approximated by a third-order polynomial
because of its embedded backward difference filter of order $d=4$
(Craigmile et al., 2004).
The wavelet variance curve for this filter also exhibits a peak
at scale $\tau_5$.
The corresponding estimated characteristic scale
is $\hat \tau_{c,5} \doteq 16.0$~months,
with an associated 95\% CI of $[11.1,23.3]$,
all of which are in reasonable agreement
with the results from the Haar wavelet.
This example points out that,
because of differencing operations embedded within wavelet filters,
there is no need to detrend a time series prior to a wavelet analysis
if care is taken in selecting a wavelet filter of appropriate
length $L_1$ to handle the nature of the apparent trend in the series.

\subsection{Coherent Structures in River Flow}\label{sec:riverflow} 
Figure~\ref{fig:figFiveTimeSeries}(b) shows 
a time series capturing so-called `coherent structures' (such as boils or eddies)
in river flows
(Chickadel et al., 2009; data courtesy of Alex Horner-Devine and Bronwyn Hayworth,
Department of Civil and Environmental Engineering, University of Washington).
The 2048 values shown in the plot
are from a longer series of length $N=29972$
that has a sampling interval of $\Delta = 1/25$~sec
and spans a little less than 20~min
(the subseries in the plot is from the first 82~sec).
This time series is derived from measurements from three transducers
and a velocity profiler set on the bottom of the
Snohomish River Estuary in Washington State
immediately downstream of a sill pointing upwards.
The structures are essentially quasi-periodic upwellings from the river
that appear as temporary `blobs' on the surface of the river.
Each blob dissipates within a second or so,
and then another blob forms sometimes later.
As the tide increases,
the water velocity increases,
and the frequency at which the blobs occur
appears to increase.
 
Videos of the river surface clearly show these boils qualitatively,
but quantifying this little-understood phenomenon
using standard Fourier-based spectral analysis is problematic
because it appears as a small perturbation in a low-frequency rolloff.
By contrast,
the scale-based analysis afforded by the wavelet variance (Fig.~\ref{fig:FourTimeSeriesWVcurves}(b))
clearly displays a peak in its decomposition of the sample variance,
rendering the phenomenon as interpretable in terms of a characteristic scale.
The estimated standardized characteristic scale is $\hat \tau_{c,6} \doteq 41.1,$
which corresponds to a physical characteristic scale
of $\hat \tau_{c,6}\,\Delta =1.6$~sec,
with an associated 95\% CI of
$[1.4,1.9]$~sec.
The time-evolving properties of the boils can be studied
by estimating characteristic scales
for time series spanning successive 20-minute time intervals.
\subsection{Madden--Julian Oscillation (MJO)}\label{sec:ThirdSeries}
Figure~\ref{fig:figFiveTimeSeries}(c) shows
the first 2048 values from  
a time series of 200-hPa velocity potential anomalies
equatorward of latitude $30^\circ$N and at longitude $80^\circ$E 
(one of a number of daily MJO indices available from
{\tt http://www.cpc.ncep.noaa.gov/}).
The entire series has 2354 values
covering 3 January 1978 to 29 March 2010
with a sampling interval of $\Delta = 5$~days.
This series is one manifestation of the MJO,
which Madden and Julian~(1994) define as a 40--50 day oscillation
appearing in various atmospheric time series collected in the tropics.
The periods associated with the MJO have been revised since 1994
based upon subsequent analysis of additional time series --
the MJO is now sometimes called a 30--60~day or intraseasonal oscillation.
The wavelet variance plot for the velocity potential anomalies
(Fig.~\ref{fig:FourTimeSeriesWVcurves}(c))
has a local peak at scale $\tau_2$,
with associated standardized local characteristic scale
$\hat \tau_{c,2} \doteq 2.53$,
which converts into a physical scale of $\hat \tau_{c,2}\,\Delta\doteq 12.7$~days.
The associated 95\% CI is $[11.9,13.5]$~days.
Since a physical scale of $\tau\,\Delta$ is associated
with the interval of periods $[2\tau\,\Delta,4\tau\,\Delta]$,
the point estimate $\hat \tau_{c,2}\,\Delta$
matches up with 25--51~day oscillations
and hence with the description of the MJO as a 30--60~day oscillation.
A difficulty with using Fourier-based spectral analysis
to deduce the MJO is the lack of a standard way to determine
the beginning and end of the frequency interval associated
with this broad-band oscillation.
The notion of a characteristic scale bypasses this difficulty
and opens up a means of objectively tracking
how the MJO varies across time and over different time series.

In addition to the peak at $\tau_2$,
there is a second one at $\tau_7$,
which leads to an estimated physical characteristic scale
of $\hat \tau_{c,7}\,\Delta \doteq 304$~days
and an associated 95\% CI of $[79,1170]$.
The interval of periods associated with $\hat \tau_{c,7}$ is $[609,1217]$~days,
so this local characteristic scale suggests an oscillation
spanning two to three years
that is about 5 times weaker than the MJO.
The presence of this weak additional oscillation
is conditional upon the peak pattern
in the wavelet variance estimates being correct.
As an example of the reality check described
at the end of Section~\ref{sec:CSEstTheory},
we generated $100,000$ realization from a trivariate normal distribution
with mean $[\hat \nu^2_6, \hat \nu^2_7, \hat \nu^2_8 ]^T$
and covariance dictated by Equation~(\ref{eq:CovApproxOne}),
with $A_{j,k}$ estimated per Equation~(\ref{eq:AjkEst}).
Of these realizations,
60\% obeyed the observed $\hat \nu^2_6 \le \hat \nu^2_7 \ge \hat \nu^2_8$ pattern,
but the remaining 40\% did not,
casting considerable doubt on the validity of the observed peak pattern.
(A similar test on the MJO peak at $\tau_2$ yielded
$99,916$ realizations with the observed peak pattern
and only $84$ without.)
\subsection{Medium Multiyear Arctic Sea Ice}\label{sec:icetype} 
Figure~\ref{fig:figFiveTimeSeries}(d$'$)
shows 2048 measurements of ice thickness taken
at 1~m spacings along a transect near the North Pole in April of 1991.
The entire set of data consists of $N=49,998$ measurements extending over 50~km
and was collected by a U.S.~Naval submarine with an upward-looking sonar
(the data are archived by the National Snow and Ice Data Center at {\tt http://nsidc.org/}).
We can regard these data as a time series with a spacing of $\Delta =1$~m,
where here `time' is a surrogate for distance 
along the submarine's path under the ice
(the submarine was moving in the same direction
at a constant speed as much as possible,
and the data were recorded at regular intervals of time).

Researchers classify sea ice by thickness,
with different ice types thought to be driven by different physical processes
(Flato, 1995, and World Meteorological Organization, 2007).
One such type is called medium multiyear ice
and has a thickness from 2 to~5~m.
The horizontal dashed lines on Fig.~\ref{fig:figFiveTimeSeries}(d$'$)
demark this ice type.
Figure~\ref{fig:figFiveTimeSeries}(d)
shows a binary-valued times series
indicating the absence or presence (using 0 or 1)
of this ice type.
Figure~\ref{fig:FourTimeSeriesWVcurves}(d)
shows the Haar wavelet variance curve for this indicator series.
The curve exhibits a single broad peak
at scale $\tau_7$,
leading to an characteristic scale is $\hat \tau_{c,7} \doteq 48.9$~m,
with an associated 95\% CI of $[29.6,80.7]$~m.
We can regard this characteristic scale
as an indicator of the `typical' extent of medium multiyear ice.
In the face of other evidence that
the Arctic climate is dramatically changing, 
a question of considerable geophysical interest is
how stable the characteristic scales for different ice types
are both spatially and temporally.
Because submarines have collected data on sea-ice thickness
throughout the Arctic region since 1958,
it is possible to look at temporal and spatial variations
in estimated characteristic scales
and to use the methodology developed in this paper
as one way to assess changes
in Arctic climatology over the past 50 years.

Finally we note that there is evidence of long-range dependence
in series of ice thickness measurements
(Percival et al., 2008).
This type of dependence maps over into indicator series,
which means that
the characteristic scale $\tau_D$
of Equation~(\ref{eq:decorrelation})
would be infinite for the medium multiyear ice indicators.
By contract, the wavelet-based characteristic scale
is finite  and provides a useful summary
of one aspect of the indicator series.

\section{Summary and Discussion}\label{sec:Summary}
We have proposed a new definition for the characteristic scale
of a time series that can be modeled as an intrinsically stationary process.
The definition is based upon local peaks in a plot
of the wavelet variance versus scale.
Since the wavelet variance provides a scale-based decomposition
of the process variance,
a characteristic scale corresponds to one
that is contributing more to the overall variance
than scales surrounding it.
This wavelet-based definition of characteristic scale
has certain advantages over other definitions,
including abilities to (1)~focus on localized properties
of the process rather than asymptotic decay rates
of autocorrelation sequences,
(2)~handle certain nonstationary processes
and (3)~handle series with trends that are well approximated
by a low-order polynomial.
We have developed a large-sample theory
for an estimator of the wavelet-based characteristic scale,
and we have demonstrated the use of this theory
through Monte Carlo experiments and applications
to four representative real-world time series.

There are several avenues of research
that are outside the scope of this article,
of which we mention three of particular interest.
First, the basis for our large-sample statistical theory
for the characteristic scale estimator
is that the underlying wavelet coefficient processes are Gaussian.
This assumption does not automatically rule out the usefulness
of our theory for non-Gaussian processes.
The filtering that is required to generate wavelet coefficients
produces a central limit effect.
Thus, even if a process is non-Gaussian,
its associated wavelet coefficient processes
might be well approximated as Gaussian, particularly at large scales.
The indicator series for medium multiyear Arctic sea ice
considered in Section~\ref{sec:icetype}
is an example of a non-Gaussian series
whose large-scale wavelet coefficients
are markedly closer in distribution to Gaussian
than the original series.
While limited tests to date indicate
that the Gaussian-based large-sample theory
for $\hat \tau_{c,j}$ is reasonably valid for indicator series,
there is certainly room for additional research
that examines the question of non-Gaussianity more thoroughly.

Second, we have assumed our time series to be regularly sampled, 
but irregularly sampled series often occur in practice.
The simplest form of irregularity is missing observations
in what would otherwise be a regularly sampled series.
Mondal and Percival~(2010)
present statistical theory for an estimator of the wavelet variance
that works with gappy time series.
This theory can presumably be used
as the basis for a characteristic scale estimator
for gappy time series.
A more serious challenge is to provide a corresponding
theory for time series sampled at irregular patterns.

A third avenue for additional research
is to handle two-dimensional data (collected on a regular grid)
that display a degree of characteristic bumpiness.
Gilgai patterns in the Bland Plain of New South Wales, Australia,
provide an example of this type of data.
Milne et al.~(2010)
have analyzed these patterns
using a two-dimensional version of the wavelet variance.
The notion of a characteristic scale
for a two-dimensional isotropic field
could be the basis
for an interesting complementary analysis of these data.

\section*{Acknowledgments}
This research was supported by U.S.~National
Science Foundation Grant No.~ARC~0529955.
Any opinions, findings and conclusions or recommendations
expressed in this paper are those of the authors
and do not necessarily reflect the views of
the National Science Foundation. 
The authors thank Yanling Yu for discussions on the Arctic sea ice example.
\section*{References}

\begin{description}

\item 
Beran, J.~(1994)
{\it Statistics for Long-Memory Processes}.
New York: Chapman \& Hall.

\item 
Chickadel, C.~C.,
Horner-Devine, A.~R.,
Talke, S.~A.~and
Jessup, A.~T.~(2009)
Vertical boil propagation from a submerged estuarine sill.
{\it Geophysical Research Letters},
{\bf 36},
L10601, doi:10.1029/2009GL037278.

\item 
Cordes, J.~M.~(1988)
Space velocities of radio pulsars from interstellar scintillations.
{\it Astrophysical Journal},
{\bf 311},
183--196.

\item 
Cox, D.~R.~and Stuart, A.~(1955)
Some quick sign tests for trend in location and dispersion.
{\it Biometrika},
{\bf 42},
80--95.

\item 
Craigmile, P.~F.~(2003)
Simulating a class of stationary Gaussian
processes using the Davies--Harte algorithm, with application to
long memory processes.
{\it Journal of Time Series Analysis},
{\bf 24},
505--511.

\item 
Craigmile, P.~F., Guttorp, P.~and Percival, D.~B.~(2004)
Trend assessment in a long memory dependence model using the discrete wavelet transform.
{\it Environmetrics},
{\bf 15},
313--335.

\item 
Craigmile, P.~F.~and Percival, D.~B.~(2005)
Asymptotic decorrelation of between-scale wavelet coefficients.
{\it IEEE Transactions on Information Theory},
{\bf 51}, 
1039--1048.

\item 
Davies, R.~B.~and Harte, D.~S.~(1987)
Tests for Hurst effect.
{\it Biometrika},
{\bf 74}, 95--101.

\item 
Daubechies, I.~(1992)
{\it Ten Lectures on Wavelets}.
Philadelphia: SIAM.

\item 
Elsner, J.~B.~and Tsonis, A.~A.~(1996)
{\it Singular Spectrum Analysis: A New Tool in Time Series Analysis}.
New York: Plenum.

\item 
Flandrin, P.~(1999)
{\it Time-Frequency/Time-Scale Analysis}.
San Diego: Academic Press.

\item 
Flato, G.~M.~(1995)
Spatial and temporal variability of Arctic ice thickness.
{\it Annals of Glaciology},
{\bf 21},
323--329.

\item 
Higuchi, T.~(1988)
Approach to an irregular time series on the basis of the fractal theory.
{\it Physica D},
{\bf 31}, 277--283.

\item 
Hosking, J.~R.~M.~(1981)
Fractional differencing.
{\it Biometrika}, 
{\bf 68}, 165--76.

\item 
Isserlis, L.~(1918)
On a formula for the product-moment coefficient of any order
of a normal frequency distribution in any number of variables.
{\it Biometrika},
{\bf 12}, 134--9.

\item 
Kay, S.~M.,~(1981)
Efficient generation of colored noise.
{\it Proceedings of the IEEE},
{\bf 69}, 480--481.

\item  
Madden, R.~A.~and Julian, P.~R.~(1994)
Observations of the 40--50-day tropical oscillation -- a review.
{\it Monthly Weather Review},
{\bf 122},
814--837.

\item  
Milne, A.~E., Webster, R.~and Lark, R.~M.~(2010)
Spectral and wavelet analysis of gilgai patterns from air photography.
{\it Australian Journal of Soil Research},
{\bf 48},
309--325.

\item  
Mondal, D.~(2007)
{\it Wavelet Variance Analysis for Time Series and Random Fields}.
PhD dissertation,
Department of Statistics,
University of Washington.

\item  
Mondal, D.~and Percival, D.~B.~(2010)
Wavelet variance analysis for gappy time series.
{\it Annals of the Institute of Statistical Mathematics},
{\bf 62}, 943--966.

\item 
Nason, G.~P.~(2008)
{\it Wavelet Methods in Statistics with R}.
New York: Springer.

\item 
Percival, D.~B.~(1995)
On estimation of the wavelet variance.
{\it Biometrika},
{\bf 82}, 619--631.

\item 
Percival, D.~B.,
Rothrock, D.~A.,
Thorndike, A.~S.~and
Gneiting, T.~(2008)
The variance of mean sea-ice thickness: effect of long-range dependence.
{\it Journal of Geophysical Research -- Oceans},
{\bf 113},
C01004, doi:10.1029/2007JC004391.

\item 
Percival, D.~B.~and Walden, A.~T.~(2000)
{\it Wavelet Methods for Time Series Analysis}.
Cambridge: Cambridge University Press.


\item 
Priestley, M.~B.~(1981)
{\it Spectral Analysis and Time Series}.
London: Academic Press.

\item 
Serroukh, A., Walden,~A.~T.~and Percival, D.~B.~(2000)
Statistical properties and uses of the wavelet variance estimator for the scale
analysis of time series.
{\it Journal of the American Statistical Association},
{\bf 95}, 184--196.

\item 
Simonetti, J.~H., Cordes, J.~M.~and Heeschen, D.~S.~(1985)
Flicker of extragalactic radio sources at two frequencies.
{\it Astrophysical Journal},
{\bf 296}, 46--59.

\item 
Tsonis, A.~A., Roebber, P.~J.~and Elsner, J.~B.~(1998)
A characteristic time scale in the global temperature record.
{\it Geophysical Research Letters},
{\bf 25}, 2821--2823.

\item 
von Storch, H.~and Zwiers, F.~W.~(1999)
{\it Statistical Analysis in Climate Research}.
Cambridge: Cambridge University Press.

\item 
World Meteorological Organization~(2007)
{\it Sea-Ice Information Services in the World}, 3rd edn.
Publication 574.
Geneva: World Meteorological Organization. 

\item 
Yaglom, A.~M. (1958)
Correlation theory of processes with random stationary $n$th increments.
{\it American Mathematical Society Translations \rm (Series 2)},
{\bf 8}, 87--141.

\end{description}
\end{document}